\documentclass[runningheads]{llncs}

\usepackage{xurl}
\usepackage{hyperref}

\usepackage[T1]{fontenc}
\usepackage{graphicx}
\usepackage{color}

\usepackage{msc}
\usepackage{amssymb} 
\usepackage{amsmath}
\usepackage{chronology}

\begin{document}
\title{``We are currently clean on OPSEC'': \\ Why JD Can't Encrypt}
\titlerunning{Why JD Can't Encrypt}
%
\author{Maurice Chiodo\inst{1} 
\and
Toni Erskine \inst{2}
\and
Dennis M\"{u}ller \inst{3} 
\and
James G. Wright\inst{4}
}
\authorrunning{M. Chiodo et al.}
%
\institute{Centre for the Study of Existential Risk, University of Cambridge, 16 Mill Lane, Cambridge, CB2 1SB, United Kingdom. \email{mcc56@cam.ac.uk} (\textit{corresponding author})\and
Department of International Relations, The Australian National University, 
130 Garran Road, Acton, ACT, 2601, Australia. \email{toni.erskine@anu.edu.au}
\and 
Institute of Mathematics Education, University of Cologne, Herbert-Lewin-Str. 10, 50931 Cologne, Germany. \email{dennis.mueller@uni-koeln.de} \and
School of Computing \& Communications, Lancaster University Leipzig, Strohsack Passage, Nikolaistrasse 10, 04109 Leipzig, Germany.  \email{j.wright18@lancaster.ac.uk}
\\ \today}
\maketitle              
\begin{abstract}
We analyse the 2025 Signalgate leak of sensitive US military information by the Trump administration, addressing why confidentiality was violated (messages leaked to the press) \emph{in spite of} encryption (Signal), to deepen the socio-technical considerations when designing and deploying encryption. First, we use $\pi$-calculus to formally model the boutique secure facility setup requested by the US Defence Secretary to prove that a leak would not be prevented. We then examine how using a secure channel might still not give overall information security, as, in this case, power imbalances between personnel and officials led to the application of cryptography that compromised their operational security. We look at how cryptographic tools may have instilled a false sense of security, and led officials to ``overshare''. We then apply this analysis to the Trump administration's general desire to burn through political, legal, and now technical process, and  demonstrate  geopolitical harms that may arise from such ineffective use of cryptography in a brief use case. We conclude that, even with advancements in usability of cryptographic tools, genuine message security is still out of reach of the ``average user''.

\end{abstract}
\section{Introduction}

\subsection{Framing of the problem: why JD can't encrypt}

This paper explores the 2025 ``Signalgate'' data leak\cite{goldberg2025trump}. It aims to explain why and how, despite the presence of a usable, mathematically secure, encrypted chat app, a conversation between high-ranking US officials as far up as Vice President JD Vance still ended up on the front page of \textit{The New York Times}\cite{NYT24.03.2025}.

Our approach is motivated by the 1999 seminal paper ``Why Johnny can't encrypt''\cite{271563}, whose socio-technical analysis showed why the average user struggled with early cryptographic software. Implementations often required knowledge about the management of cryptographic keys, overloading the average user. While \textit{mathematically} secure, they were complex and lacked user-centric design principles that meant the average user usually undermined the channel's confidentiality property. In other words: the average ``Johnny'' could not encrypt.

Now the problem of why Johnny can't encrypt has been mostly solved; cryptographic tools such as Signal are more user friendly. They are easy to install, and intuitive to use for most  users, without needing expert knowledge. However, despite this technical progress, there remain socio-technical reasons why the average user still cannot make encryption ``stick'': making use of the tool does not equate to maintaining confidentiality. In Signalgate, the tool operated as designed: sending messages and inviting new participants to a chat were quick and frictionless, resulting in a journalist being inadvertently added to a chat group of government officials including JD Vance. We argue this was a foreseeable oversight, and that the state of encrypted communications is still such that reliably securing messages remains out of reach of most users. In other words: the average ``JD'' cannot encrypt \textit{sufficiently well} to keep messages confidential.

\subsection{Summary of they key events in Signalgate}\label{s:summary}

Around 13 March 2025, US National Security advisor Mike Waltz created a Signal chat group -- `Houthi PC small group' -- containing 18 high-ranking US government officials including US Vice President JD Vance and Secretary of Defense Pete Hegseth\cite{Watson.29.03.2026}, but inadvertently added \textit{The Atlantic} editor Jeffrey Goldberg. At the time, Hegseth had a SCIF (Sensitive Compartmented Information Facility) at his residence; a typical room-sized `secure box' for sending and receiving secure communications. However, Hegseth had a boutique setup in his SCIF; a ``unique system [...] designed to mirror and access the contents of the Secretary’s personal cell phone from in his Pentagon office, connecting a keyboard, mouse, and monitor by cable to his personal cell phone, which was located outside of his office.''\cite[p.~40]{IsG}, as personal phones were forbidden within it. On 14 March 2025, Hegseth received an email (in his boutique SCIF) detailing upcoming US air and missile military strikes on Houthi forces in Yemen, to happen the following day\cite[p.~i]{IsG}. On March 15 2025, 2 hours before the strikes were due to begin, Hegseth posted specific details from the aforementioned email into the group chat\cite[p.~i]{IsG}. On 24 and 26 March 2025, Goldberg published details, including transcripts, from the group chat in \textit{The Atlantic}\cite[p.~i]{IsG}.

\subsection{The continued difficulties of encrypting}

We argue that secure communication should fulfil the following two notions:
\begin{enumerate}
    \item \textbf{Technically secure:} The ability of a sender to mathematically prove the confidentiality of their message when transmitted over an insecure channel in such a way that no eavesdropper can intercept and read it.
    \item \textbf{Socio-politically secure:} The ability of a sender to keep their message off the front page of \textit{The New York Times} for at least 2 weeks\footnote{An arbitrarily-chosen duration, simply to illustrate the point.} after sending. 
\end{enumerate}

In Signalgate, JD and his colleagues were able to be technically secure, but not socio-politically secure. And the difference between these is what we analyse. Lowry et al.\cite[p.~16]{lowry2025signalgate} describe this as the difference between securing data at rest, and data in transit. But there is more nuance here; Goldberg did not decrypt data in transit, nor did he reach in to access data at rest in an unauthorised way. He was simply given the data, albeit through a convoluted sequence of events.

Our analysis is not a direct critique of the Signalgate participants. Rather, it raises awareness of the continued difficulty for non-technical and non-expert people to secure their communications, even with usable, frictionless cryptographic software. That is: why JD can't encrypt. Our main message is simple: just because users can now operate an encrypted chat app and other new technologies, still does not mean they can understand and control all the surrounding processes and circumstances sufficiently well to ensure their communications \textit{remain} secure. This persisting lack of knowledge, combined with competing or additional interests beyond confidentiality, might result in abuse of cryptographic tools that are outside the scope of their traditional threat models. And the consequences can be severe if such cryptography is (mis)used in high-stakes scenarios. Thus, Signalgate is a (telling) example of a wider problem: Johnny may well be able to encrypt now, but JD still cannot.

\subsection{Outline of the paper}

Our paper is structured to cover technical, then socio-technical, then geopolitical, considerations in succession. First, Section \ref{s: model} provides a formal analysis using $\pi$-calculus of Hegseth's SCIF setup during Signalgate.  Section \ref{s:proof} then proves that such a setup was never going sufficiently constrain the actions of various actors, as the infosec the Department of Defense (DoD) had hoped it would. Section \ref{s:discussion} builds on these, with the following socio-technical and geopolitical analysis: 

Section \ref{s:overall} analyses the difference between having secure channels, and overall information security.  We study several surrounding factors that must be addressed, beyond the encryption tool itself, to reliably preserve confidentiality. Viewing SCIF-building as an example of \textit{tool integration}, Section \ref{s:imbalances} covers potential power imbalances between users, and those carrying out any integration of such cryptographic tools. We show that user desires and authority can win out over secure integration practises, leading to insecure processes. Section \ref{s:false_sense} covers the false sense of security cryptographic tools can give, and how they encourage users to share \textit{more} information if they \textit{perceive} to have a secure system. 

Section \ref{s:burning} presents the  case of political leaders (i.e., the Trump administration) \textit{burning through} government technical process with cryptography, to consolidate political power. Section \ref{s:geopolitics} explores some geopolitical consequences of sensitive political or military discussions happening in insecure ways and leaking to adversaries, demonstrating that ``unfit encryption'' can be devastating both for users and wider society.

Section \ref{s:concl} concludes that \textit{genuine} confidentiality remains inaccessible to the ``average user'' on their own. And so, for high-stakes communications, encryption is still insecure without proper assistance and surrounding processes.

\section{Proposed mathematical model of the SCIF}\label{s: model}

We present here a mathematical description of Hegseth's SCIF setup given in the Inspector's General report\cite{IsG}. This encapsulates the core knowledge and abilities of a \textit{SCIF operator} (SCIF-O) and their subordinates, allowing us to formally reason about the (non)communication flows between actors in and around the SCIF. In doing so, we will mathematically show how Hegseth's boutique SCIF setup was destined to fail, as the SCIF did not restrict \textbf{earnest} abusive actions.

Practitioners of security formal verification and cryptography ``\emph{aim at replacing the necessity for trusting human behaviour by mathematically enforcing and disabling specific behaviour that the systems' designers deem to be (un)-desirable}''\cite{mistrust}. Traditionally one seeks to prevent adversaries from \textbf{deceiving} honest actors into sending them a secret message or encryption key by reordering messages. So unaccountability is formulated as repudiating\cite{ZHOU1997267} (lying about) having received a message as a means of furthering the deception needed to be sent the secret message. However, we take a different approach using applied \textit{$\pi$-calculus}, which builds upon the tradition  (from Bella et al.\cite{10.1007/11542322_2,10.1007/978-3-642-30436-1_23}) of seeing how verified secure protocols can be utilised by abusive actors who \emph{have no interest in deception}, but are \textbf{earnestly \& openly} undermining the SCIF's security properties. It analyses adversaries who \emph{utilise cryptography to undermine an authorised centralising actor}\cite{mistrust} \textit{from being able to know and order events}. The formalism encapsulates \textit{who} the actors are, \textit{what} communication channels exist, \textit{which} actors have knowledge or visibility of these channels, and \textit{how} messages move around the SCIF. This allows us to formally depict what (undesirable) events can happen in such a setup.

We use the $\pi$-calculus \textit{process algebra}, as it models 1) the concurrent evolution of states of distributed actors as they transmit messages through the system, and 2) the system's communication channels  used for such messaging. The Appendix gives a review of the core elements and operations used to model SCIF processes. 

\subsection{Ordering events in the SCIF}\label{sec:SCIF}

We  use applied $\pi$-calculus to model the centralisation\cite{mistrust} of all IT communications in a SCIF, formalising what an actor can know or learn about the communications occurring within it. The goal of the SCIF-O is to ensure that all IT communications are recorded and that their causal order can be deduced. We model the record keeping with the abstraction of a \textit{vector clock} ($\mathit{VC}$): a data structure that allows distributed actors to detect causal violations in the network\cite{van2023distributed}. We assert for this model that the property of \textit{accountability} holds so long as the $\mathit{VC}$ maintains a complete record of all the subordinate clock communications. This accountability property is violated if an \textit{unauthenticated actor} communicates with someone in the SCIF, or if there are \textit{unauthorised channels} in the SCIF itself. And as this model focuses on \emph{how} unauthorised channels are created (rather than the \textit{order} of unaccountable messages), we are approximating to a $\mathit{VC}$ (rather than a \textit{realtime clock}), as we are not modelling Lamport's `Anomlous Behaviour'\cite{10.1145/359545.359563}.

Equation \ref{equati2} gives the formalised model of a $\mathit{VC}$. Note that the superscript variable $\mathrm{i}$ represents the immutable nature of an identity clock authorised to be in the SCIF by the SCIF-O, and $tr^{\mathrm{\mu, i}}$ is the transmission from $\mathrm{\mu}$ to $\mathrm{i}$.
\begin{equation}\label{equati2}
	\begin{split}
	    SO \equiv \nu\, \mathit{VC}^{\mathrm{i}}_{init} . cm (\mathit{VC}^{\mathrm{i}}_{init})\ |\ ( cc <\mathit{VC}^{\mathrm{i}}_{\alpha}> . \mathit{IncEle}(\mathit{VC}^{\mathrm{i}}_{\alpha},\mathrm{i}) .  cm (\mathit{VC}^{\mathrm{i}}_{\alpha}) )!\ |\\
        ( cm <\mathit{VC}^{\mathrm{i}}_{m}> .  ( cc (\mathit{VC}^{\mathrm{i}}_{m}) + cs (\mathit{VC}^{\mathrm{i}}_{m}) + cr (\mathit{VC}^{\mathrm{i}}_{m}) ) )!\ |\\
        (cs <\mathit{VC}^{\mathrm{i}}_o> . \mathit{IncEle}(\mathit{VC}^{\mathrm{i}}_{o},\mathrm{i}) . ( cm (\mathit{VC}^{\mathrm{i}}_{o}) | tr^{\mathrm{i,\mu}} (\mathit{VC}^{\mathrm{i}}_{o}) ) ) !\ | \\
        ( ( tr^{\mathrm{\mu, i}} <\mathit{VC}^{\mathrm{\mu}}_{r}> | cr (\mathit{VC}^{\mathrm{i}}_{s}) . \mathit{IncEle}(\mathit{VC}^{\mathrm{i}}_{s},\mathrm{i})) . \mathit{MaxVec}( \mathit{VC}^{\mathrm{\mu}}_{r}, \mathit{VC}^{\mathrm{i}}_{s}) . cm (\mathit{VC}^{\mathrm{i}}_{s} ) )!  
	\end{split}
\end{equation}
The concurrent processes in Equation \ref{equati2} can be described, in listed order, as:
\begin{enumerate}
    \item \textit{Intitialisation}: This creates the list $\mathit{VC}$ of the number of clocks in  the SCIF. Each list member keeps track of how many events have happened at the clock with that index, and is sent with every message (see\cite{van2023distributed} for details).
    \item \textit{Internal event}: This signifies that an intra-actor non-communicative event that needs to be recorded as having occurred. 
    \item \textit{Memory}: Each iteration of this process takes in the current $\mathit{VC}$ list and then forwards it to the next process that needs the current copy, via either the channels $cc$ (internal action), $cs$ (transmitting) or $cr$ (receiving).
    \item \textit{Send}: Denotes when $\mathit{VC}$ takes the List (Equation \ref{equati}) from memory, increments its index by one, and  sends copies of it to the recipient and to memory.
    \item \textit{Receive}: When the List is received, the local $\mathit{VC}$ increments it's index by one, compares the two $\mathit{VC}$ lists, and takes the largest value from either list.
\end{enumerate}

\subsection{Actors within the SCIF communication system}

We now formalise the other actors within and around Hegseth's SCIF to model the evolution of their communications during Signalgate, and present the additional processes that they were required to have to prevent the SCIF-O from learning and recording communications within the SCIF. Ultimately, these allowed for this distributed system to evolve into the violations of accountability. These actors include the SCIF operator, the honest actor Alice, JD Vance ($\mathit{JD}$) and Pete Hegseth ($\mathit{H}$), \textit{The Atlantic} journalist, and Signal's recipient discovery mechanism. From hereon, we say that a SCIF actor is \textit{authenticated} (abbreviated \textit{auth-n}) if they are on the SCIF-O's list, and if they only communicate via over authorised channels to other auth-n actors then we say they are \text{auth-z}. We write that an actor is \textit{auth-(n/z)} if they are both auth-n and auth-z.
\\\mbox{}
\\\textbf{SCIF-O:} Denoted $SO$, this is the centralised SCIF process to keep track of all events, and their order, to ensure \textit{accountability}. They are just the vector clock. 
\\\mbox{}
\\\textbf{``Honest Actor'' (Alice):}
 Denoted by $A$, this is an auth-(n/z) actor that the SCIF-O has prescribed a specific set of channels and other actors they can communicate with. They are a $\mathrm{VC}$ process, with an additional process that sends their $\mathrm{VC}$ list to the SCIF-O. Equation \ref{honest} below gives their full model.  
 
\begin{equation}\label{honest}
	\begin{split}
	    A \equiv \nu\, \mathit{VC}^{\mathrm{i}}_{init} . cm (\mathit{VC}^{\mathrm{i}}_{init})\  |\ ( cc <\mathit{VC}^{\mathrm{i}}_{\alpha}> . \mathit{IncEle}(\mathit{VC}^{\mathrm{i}}_{\alpha},\mathrm{i}) .  cm (\mathit{VC}^{\mathrm{i}}_{\alpha}) )!\ |\\
        ( cm <\mathit{VC}^{\mathrm{i}}_{m}> .  cuc (\mathit{VC}^{\mathrm{i}}_m) . ( cc (\mathit{VC}^{\mathrm{i}}_{m}) + cs (\mathit{VC}^{\mathrm{i}}_{m}) + cr (\mathit{VC}^{\mathrm{i}}_{m}) ) )!\ |\\
        ( (cs <\mathit{VC}^{\mathrm{i}}_o> | \nu\, sm) . \mathit{IncEle}(\mathit{VC}^{\mathrm{i}}_{o},\mathrm{i}) . ( cm (\mathit{VC}^{\mathrm{i}}_{o}) | tr^{i'\mu} (sm,\mathit{VC}^{\mathrm{i}}_{o}) ) ) !\ | \\
        ( ( tr^{\mathrm{\mu' i}} <m,\mathit{VC}^{\mathrm{\mu}}_{r}> | cr (\mathit{VC}^{\mathrm{i}}_{s}) . \mathit{IncEle}(\mathit{VC}^{\mathrm{i}}_{s},\mathrm{i})) . \mathit{MaxVec}( \mathit{VC}^{\mathrm{\mu}}_{r}, \mathit{VC}^{\mathrm{i}}_{s}) . cm (\mathit{VC}^{\mathrm{i}}_{s} ) )!\ | \\
        (cuc <\mathit{VC}^{\mathrm{i}}_u> . tr^{\mathrm{i,SCIF-O}} (\mathit{VC}^{\mathrm{i}}_u)) !
	\end{split}
\end{equation}
where here the \textit{additional} processes for Alice (on top of the VC processes) are:
\begin{itemize}
    \item \textit{Memory}(3): Every $\mathit{VC}$ list update is sent via name $cuc$ to the process forwarding it to $SO$ (assumes  $SO$ can process/receive every event in real time).
     \item \textit{Send}(4): Generates a secret message alongside calling the current $\mathit{VC}$ from memory, before sending them both in polyadic channel $tr^{\mathrm{i,\mu}} (sm,\mathit{VC}^{\mathrm{i}}_{o})$.
     \item \textit{Receive}(5): Alice can now receive a message and list in a polyadic channel.
     \item \textit{Forward to $SO$}(6): Each $\mathit{VC}$ list update is sent to this process and then sent to $SO$ via name $tr^{\mathrm{i,SCIF-O}} (\mathit{VC}^{\mathrm{i}}_u)$.
 \end{itemize}
This describes the \textit{honest actor}, who communicates only on authorised channels and only to authorised actors, records the occurrence of all their communications, and transmits them to the SO to keep a full record of what was sent and when.
\\\mbox{}
\\\textbf{JD Vance and Pete Hegseth:} Denoted by $JD$ \& $H$ respectively, they are the model's adversaries,  trying to undermine the accountability property by ensuring  $SO$ does not have a complete set of events, or is unable to determine their causal order. They achieve this through violations of auth-(n/z). They are an honest actor (Equation \ref{honest}), with the additional  process abilities given  in Equation \ref{JDtown}:
\begin{equation}\label{JDtown}
    \begin{split}
        B \equiv ((cs <\mathit{VC}^{\mathrm{i}}_o> | \nu\, m) .  \mathit{IncEle}(\mathit{VC}^{\mathrm{i}}_{o},\mathrm{i}) . ( cm (\mathit{VC}^{\mathrm{i}}_{o}) | (h^{\mathrm{s,p}} (m)  ) ) !\ |\\
        ( ( (h^{\mathrm{s,p}}<sm>)| cr (\mathit{VC}^{\mathrm{i}}_{s}) . \mathit{IncEle}(\mathit{VC}^{\mathrm{i}}_{s},\mathrm{i})) . \mathit{MaxVec}( \mathit{VC}^{\mathrm{\mu}}_{r}, \mathit{VC}^{\mathrm{i}}_{s}) . cm (\mathit{VC}^{\mathrm{i}}_{s} ) )!\ |  \\
        (h^{\mathrm{s,p}}<um>)! | (h^{\mathrm{s,p}} (ue) )!   
    \end{split}
\end{equation}
This describes the bridge between the adversaries' SCIF-cleared computer and their personal  Signal app (we infer from\cite{IsG} that $SO$ is not recording $m$, which can be a message or another channel). These three additional processes are: 
\begin{enumerate}
    \item An internal SCIF process generates and sends $m$ out of the SCIF ($\mathrm{s}\rightarrow\mathrm{p}$). We model the act of sending, but not the content, being added to $\mathit{VC}$.
    \item An internal SCIF process where the SCIF device receives a message from Signal ($\mathrm{p}\rightarrow\mathrm{s}$), also recorded by the $\mathit{VC}$.
    \item An instance of Signal receiving secrets from the SCIF $(h^{\mathrm{p,s}}<um>)$, and sending messages into the SCIF $(h^{\mathrm{p,s}} (ue) )$. Does not send a $\mathit{VC}$ list to $SO$.
\end{enumerate}
\textbf{\textit{The Atlantic} Journalist:}
Denoted by $U$, they have neither auth-(n/z)  to communicate in the SCIF. Their (one) process is given in Equation \ref{eq:atlantic}:
\begin{equation}\label{eq:atlantic}
    U \equiv (a<z>)!\ |\ (\nu\, x.b(x))!
\end{equation}
This models that $U$ just receives and transmits messages on channels $a$ \& $b$. They \emph{do not} have a $\mathit{VC}$ function, as they are outside the SCIF.
\\\mbox{}
\\\textbf{Signal:} Denoted $S_b$, this represents Signal's backend, and models the recipient discovery mechanism of the application, given in Equation \ref{sig}:
\begin{equation}\label{sig}
    S_b \equiv (sigr^{\mathrm{i,sig}}<cn,r>.fu(cn,r)) !\ |\ (fu<cn,r> . sigs^{\mathrm{sig,r}} (cn))!  
\end{equation}
$S_b$ receives  channel name $cn$ and intended recipient  $r$, sends it through its backend,  then sends the new channel name to the recipient (with no $\mathit{VC}$ function).

\section{Proofs that the SCIF wouldn't ensure accountability}\label{s:proof}

Having the SCIF and actors in $\pi$-calculus, we show the SCIF's failure to prevent message leaks \emph{without the $SO$ recording it} was a possible \textit{system trace},\footnote{i.e., with the given SCIF and actor setup, one can deduce a message leak is possible, even without surprise actions by external actors or malfunctions of the setup itself.} with no security controls against \textbf{earnest} adversaries beyond them acting honestly. That is, we show that a socio-political leak was a logically-derivable consequence of Hegseth's boutique SCIF setup; its architecture could not prevent such a leak.

\subsection{Process models of violations of auth-(n/z)}

We start by formalising what it means for Vance and Hegseth to \textit{violate} the  accountability processes of the SCIF, by first presenting  examples of the processes that violate auth-(n/z). We then prove that the additional processes needed for Vance and Hegseth to reduce into these violating processes can be added without the $SO$ knowing, due to the Signal process' recipient discovery mechanism.
\\\mbox{}
\\\textbf{unauth-z:}
Let $JD$ \& $H$ be made up of the processes described in Equation \ref{JDtown}. The accountability property \text{auth-z} asserts that all communication between $JD$ \& $H$ must be recorded by $SO$. Their process take the concurrent form of Equations \ref{honest} and \ref{JDtown}, given by:
\begin{equation}
    JD \equiv A\ |\ B 
\end{equation}
and the accountability violating process is given by:
\begin{equation}
    \begin{split}
        JD | un^{\mathrm{jd,h}}<nrms>\ |\ un^{\mathrm{h,jd}}(nrmr)\\
        H | un^{\mathrm{h,jd}}<nrms>\ |\ un^{\mathrm{jd,h}}(nrmr)
    \end{split}
\end{equation}
where $nrms$ is a non-recorded message and $un$ is an unauth-z channel. 
\\\mbox{}
\\\textbf{unauth-n:}
An \textit{auth-n} violation occurs when an unauthorised actor can send or receive messages from inside the SCIF (where channel $a$ isn't known to $SO$):
\begin{equation}
    JD\ |\ a^{\mathrm{j,u}} <nrms>
\end{equation}

\subsection{Proving the violations of auth-(n/z)}\label{sec:proof}
We now formally prove the violations of the accountability properties auth-(n/z), as well as why the $SO$ could not fully reconstruct the events.
\\\mbox{}
\\\textbf{Proof of violation of auth-z:} In Signalgate,  $JD$ \& $H$ wished to communicate via a channel $SO$ doesn't know about (violating auth-z). We prove this is possible in the boutique SCIF architecture (using the  equivalences from the Appendix) by showing that through  Signal, $JD$ can create a new channel name without it being in a $\mathit{VC}$ list (i.e., without $SO$ learning of it). Line 3 of Equation \ref{JDtown} gives: 
\begin{equation}
    (h^{\mathrm{s,p}}<um>)!\ |\ (h^{\mathrm{p,s}} (ue) )!  
\end{equation}
Rules $7$ and $9$ from Table \ref{struct-con} can then be used to expand this process to: 
\begin{equation}
    (h^{\mathrm{s,p}}<um>)!\ |\ (h^{\mathrm{p,s}} (ue) )!\ |\ \nu\, un \emptyset
\end{equation}
This then becomes the following context hole:
\begin{equation}\label{eq:hole}
    (h^{\mathrm{s,p}}<um>)!\ |\ (h^{\mathrm{p,s}} (ue) )!\ |\ \nu\, un [\bullet]
\end{equation}
The context hole can be used to insert Signal ($S_b$) to generate a name with the Signal backend process described in Equation \ref{sig} as follows: 
\begin{equation}
    C_0[sigr^{\mathrm{i,sig}}<un, z>]\equiv(h^{\mathrm{s,p}}<um>)!\ |\ (h^{\mathrm{p,s}} (ue) )!\ |\ \nu\, un\; sigr^{\mathrm{i,sig}}<un,z>
\end{equation}
The creation of the unauth-z channel is structurally congruent with the secured process. That is, we have proven that $JD$ \& $H$ can create a communication channel without $SO$ learning of it, as it does not sit in the $\mathit{VC}$ list  $\square$. 
\\\mbox{}
\\\textbf{Proof of violation of auth-n:} We can also formally show that $JD$ can send and receive messages from inside the SCIF, even without authorisation by $SO$ (violating auth-n). The proof of auth-n violation follows that of auth-z,  the only difference being that the process from Signal (Equation \ref{sig}) inserted into the context hole in Equation \ref{eq:hole} is $sigr^{\mathrm{i,sig}}<a, u>$; we do not repeat it here $\square$.
\\\mbox{}
\\\textbf{A proof of why $SO$ can't reconstruct events:} Our formalised model of  Signalgate has proven that, by design, the SCIF setup could \textit{never restrict}\cite{mistrust} $JD$ \& $H$ from  violating accountability through violations of auth-(n/z). In addition, we can also  model why the Inspector's General report\cite{IsG} was \textit{unable to reconstruct} the order of events  after the fact. The concurrent Signal process also provides a mechanism for \textit{message deletion}, $S \equiv S_b\ |\ S_d$, given in Equation \ref{eq:del}:
\begin{equation}\label{eq:del}
\begin{split}
    S_d \equiv  un^\mathrm{{z,h}}<nrms> . \mathit{Delete}(nrms) | un^{\mathrm{h,z}}(nrmr) . \mathit{Delete}(nrmr)
\end{split}
\end{equation}
Here,  $Delete()$  erases the record, ensuring that the $SO$ cannot reconstruct the  order of events after the fact. So the property of accountability is lost, as one cannot guarantee  the $SO$ is able to correctly assign blame for misbehaviour\cite{10.1007/978-3-642-04444-1_10}.

\subsection{Undermining accountability}

From Section \ref{sec:SCIF}, honest actors may  only communicate when  authenticated, and only on authorised channels (i.e., when they have auth-(n/z)), for the $SO$ to track all SCIF communications. But in Hegseth's boutique SCIF, Signal could create  unauthorised (unauth-z) channels between (un)authorised ((un)auth-n) actors. Section \ref{sec:proof} formally shows that once the SCIF-Signal tunnel was created, no security controls remained to  prevent or detect Hegseth creating the unauth-z channel, because of the \emph{encryption} protecting those communications. Signal's own encryption ultimately compromised the security of the SCIF. The $SO$ was also unable to discover what was said (in technical terms: unable to completely recover synchronisation of state) because of Signal's delete function,\footnote{Again, more security within Signal led to less security in the SCIF.} so they could never know the extent to which the system's accountability property was compromised. So the security of the SCIF was compromised by  \textbf{earnest} behaviour  outside the bounds of what is usually considered ``abusive'' in formal verification models. Vance and Hegseth neither lied nor deceived, so their behaviour was not constrained by the SCIF's security controls\cite{mistrust}.

\section{Socio-technical and geopolitical analysis of Signalgate}\label{s:discussion}

\subsection{The difference between secure channels, and overall security}\label{s:overall}

Those involved in Signalgate successfully set up an encrypted channel, by installing Signal on their devices, creating a chat group, adding members, and sending messages in this chat. However, a journalist from \textit{The Atlantic} was inadvertently added, who subsequently published their messages. So why did mastering \textit{mathematical} encryption (i.e., \textit{technical encryption}) not translate to actively mastering \textit{meaningful} encryption  (i.e., \textit{socio-political encryption})?

Just because some mathematical encryption process is implemented, does not imply one has meaningful encryption and overall security.\footnote{Consider a  hypothetical situation of a user receiving an encrypted file, successfully decrypting it, and then saving it to a folder on their  drive that they had forgotten (or did not know) automatically synchronises to an unencrypted cloud storage.} Without adequate surrounding processes which the `average user' may not be aware of, small oversights can compromise overall security. The complex maintenance of and dealings with  these surrounding factors appear hidden behind Signal's ease of use; it is not something that can be fully coded into a messaging app.\footnote{A month after Signalgate, Mike Waltz was observed reading government messages on his phone in front of ``a room full of reporters'' who took pictures\cite{ferguson2025waltz}.} At present, dealing with all these requires dedicated expertise, including a deep knowledge and understanding of cyber- and information security.\footnote{Lowry et al.\cite{lowry2025signalgate} apply the NIST Cybersecurity Framework to Signalgate, providing a  detailed discussion of knowledge gaps and the necessity of expert advice.} Ensuring securing communications is not a matter of maximising successes, but of minimising failures.

To stand any chance of using a secure communications tool \textit{securely} requires its proper integration\footnote{The setting up and running of a SCIF is a form of technological \textit{integration}, and so SCIF operators are \textit{integrators} in this process; a role described in detail in\cite{Muller_Chiodo_Sienknecht_2025}.} into the existing infrastructure, as is true of any high-stakes tool\cite{Muller_Chiodo_Sienknecht_2025}. Simply taking a `crypto app' off the shelf and using it does not secure communications, even though these systems are now sufficiently user-friendly that the average user can obtain the software, install it on standard hardware, and encrypt/decrypt messages reliably (compared to 20 years ago\cite{271563}). These same `average users' are not yet able to  secure the environment surrounding their encryption software. This includes dealing with issues such as insecure devices, background software that might cause data leaks, systems being used in physically insecure locations, and inadequate methods to verify recipients. 

So what integration failures occurred with Signalgate? From\cite{IsG} and public information sources, we can deduce a few; there may have been more. Section \ref{s:summary} covered Hegseth's boutique setup, and the tunnel which allowed him to exfiltrate detailed information from the SCFI in almost realtime. But his use of Signal was what allowed that information to get out of his ``circle of trusted people'', and into \textit{The Atlantic}. Hegseth was part of a Signal chat group ``Houthi PC small group'' `built' by Donald Trump’s national security adviser Mike Waltz\cite{Lowell.04.06.2025,Watson.29.03.2026}. Waltz had also (inadvertently) added the editor of \textit{The Atlantic}, Jeffrey Goldberg, to the group. How? Back in October 2024, Goldberg had emailed the Trump campaign team about a story. The team sought help from Waltz to respond, and so the content of Goldberg's email (which included his phone number) was sent by (then) Trump spokesperson Brian Hughes to Waltz. Waltz’s iPhone then did an automatic ``contact suggestion update'', finding Goldberg's number within the message and suggesting it as a new number for \textit{Hughes} (as the sender of the message), which Waltz accidentally accepted. When Waltz went to add  Hughes to the Signal group in March 2025, he already had  Goldberg's number saved under Hughes' name, and that was the number that was added. This group consisted of ``18 Trump officials'', going as high up as the Vice President, JD Vance\cite{Watson.29.03.2026}; unfortunately, none of them seemed to notice that Goldberg had been added, even though Signal identified him in the group with initials `JG'\cite{Jacobs.29.03.2026b}. 

The integration errors here were manifold\cite{lowry2025signalgate}. Top-level officials were using a publicly-available cryptographic tool (Signal) for highly sensitive communications (military strikes). This software was used on devices with other software running (iPhone contact suggestions). No checks were done to verify phone numbers corresponded to intended recipients (Goldberg's number saved as Hughes'). No-one queried all the chat participants (`JG' sat in the group for weeks). And the White House had authorised the use of Signal\cite{Lowell.04.06.2025}, creating a belief that it was secure. Indeed, globally other government agencies also ``approved'' the use of Signal\cite{Taylor.05.05.2025}, so much so that high-ranking officials are  being targeted for information through their use of it\cite{Davis.29.03.2026,FranceschiBicchierai.2026}. And while it would be tempting to dismiss this as a simple case of ``the  human error of an outside party mistakenly being added to a message chain''\cite{Chow.2025}, human-machine setups have many different failure modes, with much more nuance than just ``human error''\cite{chiodo2026formalising}; in reality, how \textit{realistic} would it be to expect Waltz to know how to avoid such a mishap?

However, the full extent of the leak was made possible by Hegseth's request for a tunnel out of the SCIF, which gave a realtime link to an insecure device, to which he could (and did) copy sensitive information. We discuss how this boutique SCIF came about in the next section on power imbalances.

\subsection{Power imbalances}\label{s:imbalances}

Hegseth's requested tunnel  out of the SCIF  to his phone created a jump point, where data first moved from a secure to insecure environment. So how did he do this? SCIFs are designed to constrain  actors, to prevent them doing things that might lead to a security breach. For this to work, those who design, build, and run the SCIF -- the \textit{SCIF operators} -- must have the authority to create the SCIF in a certain way, and to encourage (or compel) those within it to behave in compatible ways.  However, this `ability to constrain' breaks down when:
\begin{enumerate}
    \item The SCIF users have authority over the operators, and 
    \item The SCIF users exercise that authority to change what the operators are doing, and/or how those within the SCIF can act.
\end{enumerate}

When exercised, this \textit{power imbalance}\cite{Muller_Chiodo_Sienknecht_2025} can lead to operators and their processes being subverted, compromised, or rendered otherwise ineffective. It doesn't matter how hard operators try to secure the environment around communicating actors, if those  actors do not want such restrictions, and moreover can order them changed or can simply ignore them.  As a simple example, those building cryptographic systems might implement environmental controls to prevent device screenshots when an app is open. But a user may realise screenshots have been disabled, and deliberately subverts it by pointing an external camera at the device; something out of control of the system builder. If there are no override mechanisms in place, they may even be able to re-enable screenshots in the system settings. Power and influence imbalances may enable users to ``go over''  (or around) the desires of the operators, \textit{and} if they choose to exercise this power, a well-thought-out secure communications system can be weakened or compromised\cite{lukes2021power}. This may happen because the users want to do something the system does not currently allow,  without realising \textit{why} such actions have been prohibited. What might initially be perceived as ill-intent, may actually boil down to a combination of ignorance of the importance and necessity of system components, and competing interests between their and the operators' goals.

From Section \ref{s:proof}, Signalgate was precipitated by the construction of a boutique SCIF at Hegseth's  residence: ``at the request of the Secretary, the JMA [Junior military assistant] requested and oversaw the installation of a unique capability through which the Secretary could access and control his personal cell phone from inside his secure office''\cite[p.~40]{IsG} (see Section \ref{s:summary} for details). The request was  ``consistent with DoD information security requirements, which did not allow for any cell phones to be brought into the Secretary’s suite''\cite[p.~40]{IsG}, but for all intents and purposes Hegseth had full access to his personal phone from within the SCIF, via this tunnel. While the  phone was not brought in, its  \textit{functionality} was. Even if the tether console  was air-gapped from other SCIF IT, Hegseth himself acted as a communication bridge over that gap, transcribing information from classified emails with significant levels of detail\cite[pp.~12,22--23]{IsG}; likely more than could just be remembered after leaving the SCIF. It can be reasonably assumed that the JMA understood the dangers of this setup\cite[pp.~25--26]{lowry2025signalgate}. 

Throughout this process, there was practically nothing constraining Hegseth. As  Secretary of Defense, his JMA was his military subordinate,  obliged to follow his lawful orders. While existing ``DoD policy prohibits the use of nonapproved commercially available messaging applications, such as Signal, for official business, with limited exceptions.''\cite[p.~3]{IsG}, providing exceptions for this lawfully rested with Hegseth.  The level of classification of the material was a moot point, as Hegseth had the authority within the DoD to determine classifications, which only the President could override\cite[p.~23]{IsG}. Yet, the only entity that could constrain Hegseth -- the White House -- had done the exact opposite, having  ``authorized the use of Signal, largely because there is no alternative platform to text in real time across different agencies''\cite{Lowell.04.06.2025}. That is, there was higher authority, and motivation, for using Signal in this way. And Hegseth specifically sought this convenience,  arguing ``so I could more easily receive non-official communications during the workday''\cite[p.~66]{IsG}. But why were such sensitive communications sent in this way? Likely because using Signal had given them a \textit{false sense of security}.

\subsection{False sense of security}\label{s:false_sense}

The Signalgate participants knew they had an encrypted channel, giving them a false sense of security. Hegseth even wrote ``We are currently clean on OPSEC'' in the  chat\cite[p.~12]{IsG}. They then used this to communicate sensitive information. So, did the introduction of cryptographic tools \textit{increase} data risks here?

It may be tempting for cryptographers to treat both message quantity and sensitivity as independent of the existence of cryptographic tools, falsely assuming that communicating parties will send the same messages \textit{anyway}, regardless of whether or not they have access to cryptographic tools (so the encrypted channel can be agnostic to all messages and sessions within). That is: $\frac{\partial \ \textnormal{messages}}{\partial \ \textnormal{crypto}} =0$. Under such a model, cryptographic tools become an \textit{unmitigated tool for good}; they add extra security to an existing message stream, and do not introduce additional risks. Even if the tools fail to secure messages, no additional harm is done, as the messages would have been sent unencrypted anyway. However, with increased access to usable cryptographic tools, message quantity and sensitivity also goes up: $
\frac{\partial \ \textnormal{messages}}{\partial \ \textnormal{crypto}} >0$. 
With access to stronger cryptography, people simply say more. In the knowledge that their messages are securely encrypted, the communicating parties feel comfortable increasing the information being transmitted (discussed in\cite[pp.~16--17]{lowry2025signalgate}), and its level of sensitivity.

Furthermore, this is not strictly a function of how good the cryptography \textit{is}, but rather, how good it is \textit{perceived to be}; it only takes users to \textit{believe} that they are communicating in a secure environment, to tempt them to start over-sharing information. One sees this in the Babington plot, where Mary Queen of Scots conspired over ``encrypted'' letters to have Elizabeth I assassinated; Mary was caught, and executed\cite[pp.~21--22]{dooley2013brief}. Had she not had access to what she \textit{thought} was an encrypted communication channel, she may have refrained from sending messages about the plot; messages that were decrypted and used as evidence to convict her. And so the relationship is perhaps better described by $\frac{\partial \ \textnormal{messages}}{\partial \ \textnormal{perceived crypto}} >0$, which can add significant amounts of risk, especially when the perceived strength of the cryptography is far better than its actual strength.

The Signalgate participants clearly had an inflated sense of security in the cryptography they were using. After all, it was a piece of software whose Terms of Service (ToS) opens with ``Signal messages and calls cannot be accessed by us or other third parties because they are always end-to-end encrypted, private, and secure''\cite{Signal.01.04.2026}. Perhaps their faith in it is unsurprising, and might explain why Hegseth was sure operational security was maintained\cite[p.~12]{IsG}. They may have been operating under the assumption that a system which has worked well for them in the past, will continue to work well in the future, turning a temporary solution into informal White House policy: ``as a temporary solution, the Trump White House told officials to use Signal as they had done during the transition instead of regular text-message chains''\cite{Lowell.04.06.2025}; giving them confidence from the highest levels of government. Although \textit{some} awareness existed that Signal was not fully secure, as participant Steve Witkoff refused to carry his personal phone on a trip to Russia (citing security concerns), there appears to have been a false belief among  participants that Signal was at least secure within the US\cite[p.~17]{lowry2025signalgate}; one sufficiently strong that JD Vance \textit{re-entered} the chat the day the story broke\cite{goldberg2025trump} to say ``This chat's kinda dead. Anything going on?''\cite[p.~51]{IsG}. A month later, Waltz was seen reading Signal in front of ``a room full of reporters''\cite{ferguson2025waltz}.

The participants may have assumed that Signal securing \textit{data in transit} extended to securing \textit{data at rest}\cite[p.~16]{lowry2025signalgate}, even though the ToS explicitly state ``You are responsible for keeping your device and your Signal account safe and secure''\cite{Signal.01.04.2026}. In a conflation of end-to-end encryption with having control over endpoints, the existence of unwanted participants was probably unfathomable to them, even though it was not the first time  a journalist entered a secure zone via fairly elementary means.\footnote{In 2007, journalists entered the secure ``red zone'' of the APEC summit. They used a fake motorcade with Canadian flags. Their `security detail' had ID passes printed with the words ``This is a joke''. And one was dressed as Osama bin Laden\cite{Valentish.19.01.2020}.} That messages \textit{should} be erased from all record via Signal's auto-delete feature, may have given them further impetus to over-share with perceived impunity. All the while, Goldberg took screenshots of the chat, preserving messages that had otherwise been wiped from Hegseth's phone when sought by the investigation\cite[p.~12]{IsG}. One wonders why they had so much faith in a piece of software whose ToS has an extensive disclaimer (in all-caps) that starts with ``YOU USE OUR SERVICES AT YOUR OWN RISK''\cite{Signal.01.04.2026}.
 
Ultimately, their faith in the setup's security led them to overshare so egregiously that the \textit{New York Post} ran with the front page headline ``Operation Overshare''\cite{Nava.25.03.2025}. These messages were extremely sensitive, containing exact details about the Yemen attack, including the timings of fighter jet strikes and tomahawk missile launches\cite[p.~12]{IsG}, and sent 2 hours before the strikes. We discuss potential catastrophic consequences of such oversharing in Section \ref{s:geopolitics}.

\subsection{Burning through the technical processes}\label{s:burning}

The Trump administration has always been committed to tearing down and replacing existing political structures and practices. Indeed, former White House Chief Strategist Steven Bannon in 2017 even referred to Trump’s agenda as involving no less than the ``deconstruction of the administrative state''\cite{Rucker.23.02.2017}. This has involved burning through the system of internal regulations and external agreements that Trump has argued is holding the US back politically and economically. Such an agenda has become even more overt in his second term. Domestically, in 2025, the Department of Government Efficiency (DOGE) was committed to overhauling the federal civil service. And in terms of its external relations, the administration has unilaterally withdrawn the US from dozens of international organisations\cite{Cameron.7.1.2026}, thereby eschewing multilateralism, international consensus, and coalition building; contributing to the dismantling of the post-WWII ``rules-based order''\cite{Carney}.

When it comes to established international norms, including those codified in international law, the Trump administration has distinguished itself by its willingness to disregard and override them, without the pretense of justification or excuse. In its interactions with other states, it consistently flouts expectation and established practice, proceeding according to its own conception of what is right and required. It has imposed trade tariffs on allies and adversaries alike, and embarked on military actions in Venezuela and Iran that violate the prohibition on the use of force enshrined in the United Nations (UN) Charter\cite{UnitedNations.1945}, without explanation.\footnote{Whereby states must ``refrain [...] from the threat or use of force against the territorial integrity or political independence of any state''\cite[Ch.~I, art.~2, para.~4]{UnitedNations.1945} with the explicit exceptions of ``the inherent right of individual or collective self-defence if an armed attack occurs'' or the UN Security Council endorsing the multilateral use of force when it recognises a threat to international peace and security\cite[Ch.~VII]{UnitedNations.1945}.} Established norms are sidestepped -- and even lack tacit acknowledgment in their abrogation through any attempt to justify overriding or reinterpreting them. In Trump's own words, ``I don’t need international law.'' Instead, he claimed to have just one source of constraint: ``My own morality. My own mind. It’s the only thing that can stop me.''\cite{Sanger.8.1.2026}. In sum, traditional constraints are rejected as inconsequential and rules are created anew.

The administration is now increasingly burning through technical norms. As an early example of this, DOGE upended technical norms regarding static data (i.e., what was \textit{on} computers and who \textit{had access} to it). With Signalgate they started to use non-government communication infrastructure (i.e., Signal) to upend how data moves around (i.e., how computers are \textit{used}). The administration worked around, and actively reduced, previously working safeguards (e.g., SCIFs), burning through the disciplinary ideals and professional standards of the involved technical experts. In doing so, they burned through long-standing record-keeping mechanisms via Signal's auto-delete function. Signalgate has brought light on how the administration is actively setting up a new normal, overriding or removing existing experts (e.g., by firing Inspectors General\cite{lowry2025signalgate}), thus burning through the very people put in place to help them.

Dealing with this requires more than the zero trust principles advocated for by Lowry et al.\cite{lowry2025signalgate}. The cryptography and cybersecurity communities need to account for such a new normal. In Signalgate, participants got away with more than expected, as socio-technical safety constraints like SCIFs could no longer rely on the existence, and enforcement, of suitable norms, laws, and regulations.

\subsection{War deliberations: geopolitical consequences of unfit encryption}\label{s:geopolitics}

Signalgate resulted in acute embarrassment for the Trump administration. However, one might argue that they got lucky. The compromised information was about a discrete set of military strikes as part of an ongoing campaign and, although Goldberg eventually made them public, he did so only after the attacks had taken place and, even then, with an effort to protect what he recognised to be classified information.\footnote{Goldberg's first article contained no  chat extracts\cite{goldberg2025trump}. However, when the Trump administration argued that nothing in the chat was classified, Goldberg concluded  it  in ``the public interest'' to share the extracts so that ``the sort of information that Trump advisers included in nonsecure communication channels'' could be scrutinised\cite{Goldberg.26.03.2025}.}  Under different circumstances, the consequences of such a breach could have been immeasurably worse -- even catastrophic. Most significantly, a radically different outcome could have followed if the leaks had been shared widely \textit{prior} to the strikes, if information fell \textit{directly} into an adversary's hands, or if even \textit{more} sensitive intelligence about large-scale strategies had leaked.  One might, for example, consider the possible consequences of Signalgate having occurred a year later, with some or all of these alternative conditions, in the context of the initiation of armed conflict against Iran.

Leaked information about state military operations -- whether related to precise details of a imminent attack, as in the case of Signalgate, or deliberations over future contingencies -- pose a clear threat to national security. With respect to forthcoming attacks, compromised intelligence about a state's planned military operation not only risks successfully obtaining the objectives, but also seriously jeopardises the lives of military personnel. As Goldberg and Harris\cite{Goldberg.26.03.2025} suggested in an attempt to highlight the gravity of the situation, ``[i]f this information -- particularly the exact times aircraft were taking off for Yemen -- had fallen in to the wrong hands in that crucial two-hour period, American pilots and other American personnel could have been exposed to even greater danger than they ordinarily would face.'' Such disclosures could also prompt -- and even provide justification for -- pre-emptive attacks against the state experiencing the  breach.\footnote{Although debated, a degree of anticipatory self-defense is often understood to be allowed under international law in cases of an imminent threat, and such leaked deliberations could be interpreted as evidence for such a threat (see\cite{Deeks.2015} for details).} Significantly, and most tragically, such a pre-emptive response could be generated even if the state that experienced the breach were just \textit{contemplating} possible future scenarios rather than actually planning an attack, thereby leading to the unintended initiation, or escalation, of a conflict. In short, it is conceivable that unfit encryption could lead to catastrophic geopolitical consequences.

It is also worth considering other geopolitical effects that could result from the disclosure of extremely candid deliberations intended to remain behind closed doors,\footnote{Such as the 2010 ``Cablegate'' leak of sensitive US diplomatic cables\cite{Manning}.} in the ``war room''.\footnote{Such as the 2022 ``Discord'' leaks of sensitive US DoD documents\cite{Discord}.} In Signalgate, trust in the cryptography brought together a degree of informality prompted by the chat, with a complacency that this high-level conversation was secure. Three potentially troublesome aspects of the leaked discussion were revealed in addition to details about tools and  timing of strikes: 1) that there was division within this executive group as to whether the strikes were warranted and why; 2) harsh evaluations of allies; and 3) a seemingly cavalier attitude towards the consequential decision to use force, and towards the inevitable loss of civilian life that necessarily accompanied it (highlighted by the particular use of emojis in the leaked messages). Such additional revelations, made possible by unfit encryption that was nevertheless trusted, have potentially far-reaching diplomatic, ethical, and security implications.

\section{Conclusion and recommendations}\label{s:concl}

Our analysis of Signalgate shows that, even with advancements in usability of cryptographic tools, genuine confidentiality is still out of reach of the ``average user''. Signalgate itself was a disaster waiting to happen, not because those involved acted \textit{wrongly}, but rather, because the traditional threat models don't consider \textbf{earnest} adversaries. A belief that ``cryptography makes us secure'' set off a chain reaction that led to sensitive government communications ending up on the front page of \textit{The New York Times} within 2 weeks; JD and his colleagues were simply unable to convert \textit{technical security} into \textit{socio-political security}. But none of this is surprising; how would the participants have \textit{known} that what they were doing was insecure? As we have argued, they couldn't have.

And so, to those who develop such security protocols, we recommend formalising new threat models that restrict these newly emerging abusive actors. These might be used in unexpected instances where the stakes are high, and the surrounding circumstances hazardous and insufficiently controlled. In the case of Signalgate, as we mentioned many times, more cryptography made things \textit{worse}. It is important to think not just about how these protocols \textit{should} be used, but also \textit{who} will use it; one cannot assume that it will prevent all abusive deployment and user behaviour. And although \textit{usability} is something that can be developed within the software itself, \textit{overall security}, as we have seen, is not. 

\newpage

\bibliographystyle{splncs04}
\bibliography{clocks}
\section{Appendix: Applied $\pi$-calculus background}
\textbf{Overview of applied $\pi$-calculus}: \cite{10.1145/3127586,MILNER19921,MILNER199241} give a detailed introduction to (applied) $\pi$-calculus. Here we  review the elements and operations used to model SCIF processes. 
Equation \ref{apcdef} defines the set of applied $\pi$-calculus elements involved in our models:  
\begin{equation}\label{apcdef}
	\begin{array}{lr}
        L,M,N::= &\mbox{Terms}\\
		\mathrm{i,j,k},\ldots&\mbox{Variables}\\
		a,b,c,\ldots &\mbox{Names} \\
		\mathit{f}(M_1,\ldots,M_n)&\ \ \ \mbox{Functions}\\
	\end{array}
\end{equation} 
Here, \textit{Terms} can be an infinite number of Names and/or Variables, or a finite number of Functions. \textit{Variables} are used to represent an actor's immutable identity. \textit{Names} then either represent messages, or communication channels. \textit{Functions} represent specific actor computations.




We  use the following $\pi$-calculus \textit{processes}. These represent the individual steps that can occur in the evolution of the communicating processes. Together, they describe the full operational evolution of the system. For the sake of brevity, here it is assumed that $x,y,z$ can be Names or Variables (as in Equation \ref{apcdef}).
\begin{enumerate}
	\item $P.Q$: A linear sequence of processes P and Q.
    \item $x<z>.P$: The reception of the name/variable $z$ from channel $x$. Any subsequent $z$ names/variables in the receiving process will be replaced by the name that is received in the process $\{y/z\}$.
	\item $x(y).P$: The transmission of name/variable $y$ along channel $x$.
    \item $\nu\, z\; z.P$: A restriction of the name/variable $z$, where sending or receiving along $z$ can only occur if the involved processes already knows of $z$'s existence. This operation is used to make explicit restriction of a name/variable, or to represent the generation of a fresh name/variable. 
    \item $P\ |\ P'$: Processes $P$ and $P'$ can evolve concurrently.
    \item $P+P'$: The evolution of the overall process $P$ and $P'$ can only evolve along one or the other.
    \item $P!$: The replication of $P$ to generate infinite copies of the same process, written as $P\ |\ P!$.
    \item $\{P/x\}$: The substitution process, where all instances of process $P$ are replaced with the name/variable $x$.
	\item $\emptyset$: Termination of a process.
\end{enumerate} 
P represents an arbitrary process that can coordinate sub processes as well as with other(s/ processes). Functions used in this model have an equational theory describing how they are treated in process evolution, expanding on\cite{10.1145/3127586}:
\begin{equation}\label{equati}
	\begin{split}
		\mathit{R,S,T}\; :\; List\\
        \nu\, \mathit{R}\\
        \mathit{IncEle}(\mathit{R}_{\alpha}, i) = \nu\, \mathit{R}_{\alpha, inc} . \{ \mathit{R}_{\alpha , inc} / \mathit{R}_{\alpha}\}\\ 
        \mathit{MaxVec}(\mathit{L,I}) = \nu\, \mathit{K}_{max} . \{\mathit{K}_{max}/\mathit{I}\}
	\end{split}
\end{equation}
where each of the four entries above is interpreted as follows:
\begin{enumerate}
    \item $List$: Defines the list object.
	\item $\nu\, \mathit{R}$: Is the initialisation of a list.
    \item $\mathit{IncEle}(\mathit{R}_{\alpha}, i)$: Increments an element of the list (where  $\mathit{R}_{\alpha, inc}[\mathrm{i}] := \mathit{R}_{\alpha}[\mathrm{i}] + 1$).
    \item $\mathit{MaxVec}(\mathit{L,I})$: compares all the elements of two lists of equal length (where here we have $\mathit{K}_{max}[\mathrm{i}] := max(\mathit{L}[\mathrm{i}] , \mathit{I}[\mathrm{i}])$).
\end{enumerate}

\noindent \textbf{Applied $\pi$-calculus equivalences:} (Applied) $\pi$-calculus has various (axiomatic) equivalence relations which can be used to formally prove  various processes are semantically or structurally equivalent, regardless of their actual state machines. Our work uses
structural congruence equivalence (denoted $\equiv$), which is used to formally state the kinds of modifications to a process that do not affect the evolution of the original process. The list of axiomatic modifications that do not affect the original process are given in Table \ref{struct-con} (these are standard, from\cite{sangiorgi2003pi}). Note that the ``$\emptyset$'' processes can be transformed into \textit{contexts}, denoted $[\bullet]$, which are points that explicitly indicate where \textit{new} processes can be inserted without affecting the original process.
\begin{table}
	\centering
	\begin{tabular}{l l r c r}
		\hline
		Index & Label & Axiom & & \\
		\hline
		1&$\textbf{SC-MAT}$&$[x=x]\pi.P$&$\equiv$&$\pi.P$ \\
		2&$\textbf{SC-SUM-ASSOC}$&$M_1 + (M_2+M_3)$&$\equiv$&$(M_1 + M_2)+M_3$ \\
		3&$\textbf{SC-SUM-COMM}$&$M_1+M_2$&$\equiv$&$M_2+M_1$ \\
		4&$\textbf{SC-SUM-INACT}$&$M+\emptyset$&$\equiv$&$M$ \\
		5&$\textbf{SC-COMP-ASSOC}$&$P_1\ |\ (P_2|P_3)$&$\equiv$&$(P_1|P_2)\ |\ P_3$ \\
		6&$\textbf{SC-COMP-COMM}$&$P_1\ |\ P_2$&$\equiv$&$P_2\ |\ P_1$ \\
		7&$\textbf{SC-COMP-INACT}$&$P|\emptyset$&$\equiv$&$P$ \\
		8&$\textbf{SC-RES}$&$\nu\, z\; \nu\, w \;P$&$\equiv$&$\nu\, w\; \nu\, z \;P$ \\
		9&$\textbf{SC-RES-INACT}$&$\nu\, z \;\emptyset$&$\equiv$&$\emptyset$ \\
		10&$\textbf{SC-RES-COMP}$&$\nu\, z \;P_1\ |\ P_2$&$\equiv$&\ $P_1\ |\ \nu\, z \;P_2$ 
        \ $\mathrm{if}z\notin fn(P_1)$\\
		11& $\textbf{SC-REP}$&$P!$&$\equiv$&$P\ |\ P!$ \\
		\hline
	\end{tabular}
	\caption{List of structural congruence axioms\cite{sangiorgi2003pi}}
	\label{struct-con}
\end{table}

\end{document}